\documentclass[10pt,letterpaper]{extarticle}
\usepackage[top=0.85in,left=1.75in,footskip=0.75in]{geometry}

\usepackage[utf8x]{inputenc}
\usepackage{upgreek}
\usepackage{graphicx}
\usepackage{layouts}

\usepackage{cite}

\usepackage{nameref,hyperref}

\usepackage{microtype}
\DisableLigatures[f]{encoding = *, family = * }

\raggedright
\usepackage{changepage}

\usepackage[aboveskip=1pt,labelfont=bf,labelsep=period,singlelinecheck=off]{caption}

\makeatletter
\renewcommand{\@biblabel}[1]{\quad#1.}
\makeatother

\usepackage{lastpage,fancyhdr,graphicx}
\usepackage{epstopdf}
\usepackage{color}

\definecolor{Gray}{gray}{.25}

\usepackage{graphicx}

\usepackage{sidecap}

\usepackage{wrapfig}
\usepackage[pscoord]{eso-pic}
\usepackage[fulladjust]{marginnote}
\reversemarginpar

\begin{document}
\vspace*{0.35in}

\begin{flushleft}
{\Large
\textbf\newline{Intensity Noise Optimization of a Mid-Infrared Frequency Comb Difference Frequency Generation Source}
}
\newline
\\

Vinicius Silva de Oliveira\textsuperscript{1},
Axel Ruehl\textsuperscript{1,2,3},
Piotr Mas\l{}owski\textsuperscript{4},
Ingmar Hartl\textsuperscript{1,*}
\\
\bigskip
\bf{1} Deutsches Elektronen-Synchrotron (DESY), Notkestrasse 85, 22607 Hamburg, Germany
\\
\bf{2} Leibniz University Hannover, QUEST-Leibniz-Research School, Institute for Quantum Optics, Welfengarten 1, 30167 Hannover, Germany
\\
\bf{3} Laser Zentrum Hannover e.V., Hollerithallee 8, 30419 Hannover, Germany
\\
\bf{4} Institute of Physics, Faculty of Physics, Astronomy and Informatics, Nicolaus Copernicus University in Toru\'n, ul. Grudziadzka 5, 87-100 Toru\'n, Poland

\bigskip
* ingmar.hartl@desy.de

\end{flushleft}

\section*{Abstract}
  We experimentally demonstrate in a difference-frequency generation mid-infrared frequency comb source the effect of temporal overlap between pump- and signal- pulse to the relative intensity noise (RIN) of the idler pulse. When scanning the temporal delay between our 130\,fs long signal- and pump pulses, we observe a RIN minimum with a 3\,dB width of 20\,fs delay and an RIN increase of 20\,dB in 40\,fs delay at the edges of this minimum. We also demonstrate active long-term stabilization of the mid-infrared frequency comb source to the temporal overlap setting corresponding to the lowest RIN operation point by an on-line RIN-detector and active feedback control of the pump-signal- pulse delay. This active stabilization set-up allowed us to dramatically increase the signal-to-noise ratio of mid-infrared absorption spectra.


\section{Introduction}
Coherent light sources in the mid-infrared (MIR) spectral region are important for numerous applications, since they provide access to the “molecular fingerprint” spectral range \cite{Schliesser2012, cossel2017gas}. In this wavelength range between 3\,µm and 15\,µm specific molecular transitions can be addressed, allowing for example to identify and measure the concentration of various molecules in a gas mixture \cite{foltynowicz2013, maslowski2014cavity, muraviev2018massively}. Since only few MIR broadband laser gain media exist \cite{sorokin2005ultrabroadband, mirov2013frontiers, jackson2012}, many sources rely on nonlinear frequency conversion \cite{petrov2015frequency, tittel2003mid}. Among numerous frequency conversion methods utilizing $\chi^{(2)}$ and $\chi^{(3)}$ nonlinearities in materials, difference frequency generation (DFG) is a widely used method \cite{erny2007mid, gambetta2008mid, Ruehl2012, Neely2011, Mayer2016, sobon2017high}. In the DFG method two near infrared (NIR) laser pulses, the pump and signal pulse interact in a $\chi^{(2)}$ nonlinear crystal material and generate a MIR idler pulse, conserving photon energy and momentum \cite{bloembergen1980conservation}. The DFG scheme is very popular in carrier-envelope-offset phase (CEP) sensitive applications, since it allows to construct sources with passively CEP stable output \cite{krauss2011all, Zimmermann2004}.

As an unwanted side-effect, the nonlinear processes in MIR DFG sources can produce excess noise both in intensity and phase \cite{Neely2011, Mayer2016, huber2017active, homann2013direct}. For many applications it is important to keep this excess noise as low as possible. In absorption spectroscopy for example, the relative intensity noise (RIN) of the light source can be the overall dominating noise source \cite{hobbs1997ultrasensitive}. In spectroscopic applications lower light source RIN directly results in a higher signal-to-noise ratio of the acquired spectra, allowing for shorter acquisition times and improving the detection limits \cite{ye1998ultrasensitive, foltynowicz2011Quantum}. In this contribution we show experimentally that fine control of the temporal overlap between pump- and signal- pulse in a MIR DFG frequency comb source can result in reduction of RIN by up to 30\,dB.

\section{Mid-infrared DFG source}
Our MIR frequency comb source is similar to the one described in Ref. \cite{Ruehl2012}. It is based on a saturable-absorber mode-locked 150\,MHz Yb:fiber oscillator seeding two cladding pumped Yb:fiber amplifiers, which are used for self-referencing the comb and for MIR generation, respectively. The oscillator can be optically referenced by phase-locking a comb-line near 1064\,nm to a kHz-level line width single frequency laser. The Yb:fiber amplifier for MIR generation produces 1.5\,W output pulses of  130\,fs FWHM pulse duration at a center wavelength of  1050\,nm and 25\,nm bandwidth (FWHM).

\begin{figure}[htbp]
\centering
  \includegraphics[width=\linewidth]{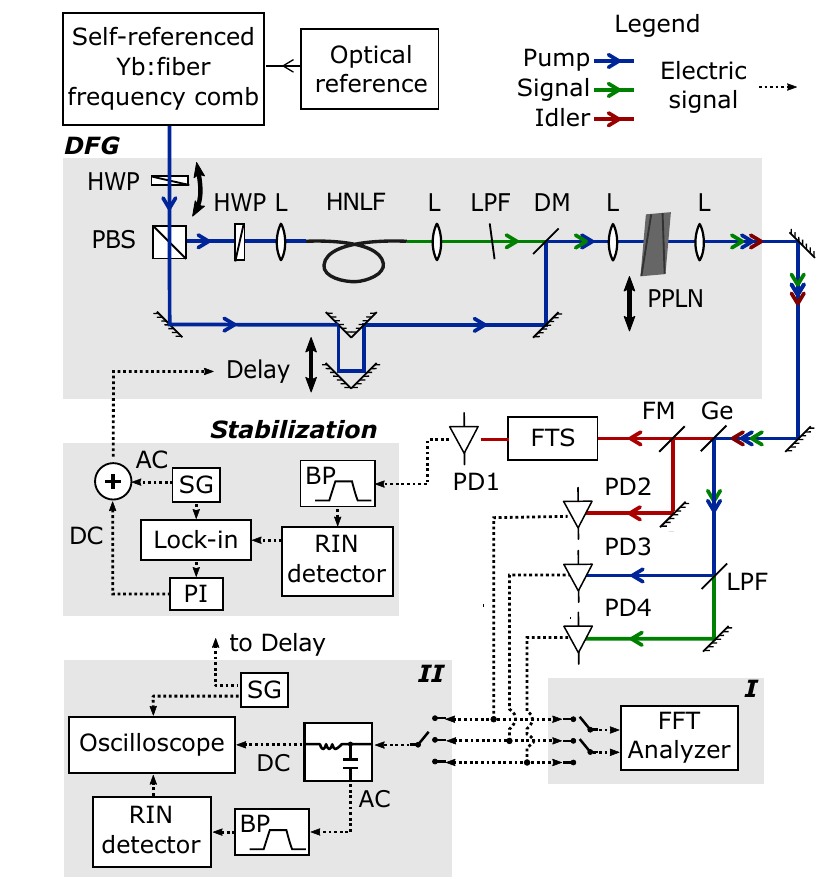}
	\caption{Schematic illustration of the experimental setup and the three main parts, labeled as: “DFG”, “Stabilization”, and setups I and II for noise characterization. HWP: half wave plates. PBS: polarizing beam splitter. L: lenses. HNLF: highly-nonlinear suspended core-fiber. LPF: long-pass filter. Ge: germanium filter. FM: flip mirror. PD1, PD2: liquid-nitrogen cooled photovoltaic InSb photodetector. FTS: home-build Fourier transform spectrometer. PD3, PD4: InGaAs PIN photodetector. BP: electronic band-pass filter. RIN detector: demodulating logarithmic amplifier. SG: signal generator. PI: proportional integral controller. The top right legend indicates color and line-shape coding to distinguish optical and electronic signal paths. Black double-sided arrows illustrate required mechanical translations for wavelength tuning.}
	\label{setup}
\end{figure}

The set-up for DFG MIR generation is schematically shown in Fig. \ref{setup}. The output of the Yb:fiber amplifier for MIR generation is divided into two parts, which are used as pump pulse and to generate the signal pulse for the subsequent DFG process. For signal pulse generation, we used the longest wavelength Raman soliton from supercontinuum generation in a highly-nonlinear suspended-core fiber \cite{Dong2008}. By controlling the coupled power into the highly-nonlinear fiber in the range between 50\,mW and 500\,mW, we can continuously control the center wavelength of the longest wavelength Raman soliton from approx. 1.2\,µm to 1.6\,µm. The output of the highly-nonlinear fiber is long-pass filtered (cut-on wavelength: 1200\,nm) and collinearly combined with the pump-pulse using a dichroic mirror. Since the longest wavelength Raman soliton is temporally separated from the remaining radiation, the long-pass filter is not required for the MIR source.
In our set-up however, the filter is beneficial for noise characterization, since it isolates the Raman soliton from supercontinuum components not involved in MIR generation.

We use a periodically poled Lithium Niobate (PPLN) crystal (10\,mm length, 1\,mm height) for difference frequency generation. The crystal has a fan-out grating, allowing to continuously adjust the poling period from 26.15\,µm to 30.65\,µm by laterally displacing the crystal. The pump- and signal- beams are spatially overlapped with 1/e$^2$ waist diameters of 180\,µm and 116\,µm, respectively in the crystal. Temporal overlap is achieved by a mechanical delay-line in the pump arm. The delay line is equipped with a piezo-mechanic transducer for fine-control.

\begin{figure}[htbp]
\centering
  \includegraphics[width=\linewidth]{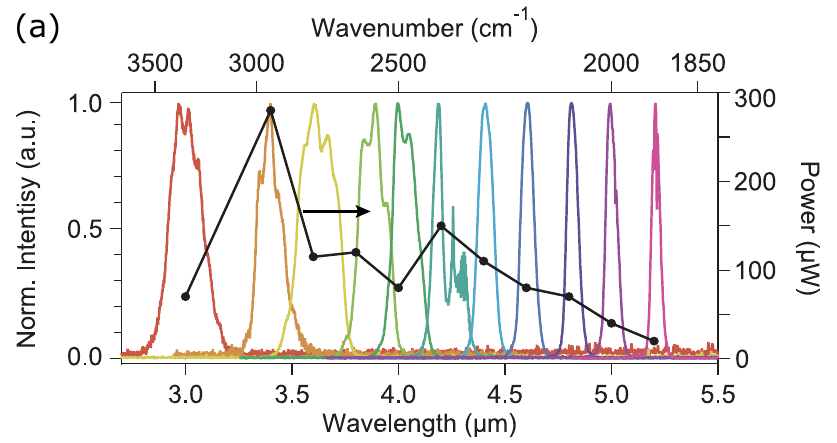} \\
  \includegraphics[width=\linewidth]{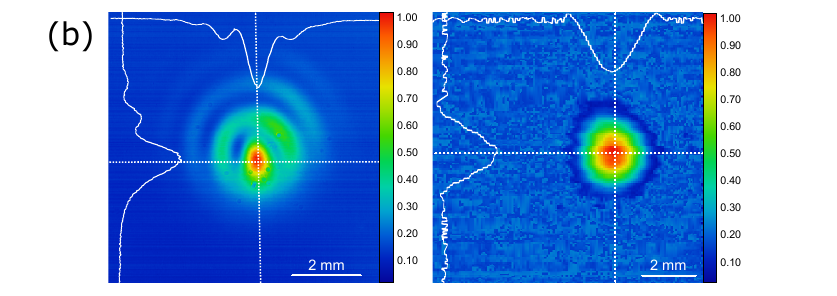}
	\caption{(a) MIR output FTS spectra and output powers at different output center wavelengths. For this measurement, the source was optimized for an absorption spectroscopy experiment at 4.6\,µm before performing the tuning test. (b) Far-field beam profiles at two different focusing conditions: left, 60\,µm beam diameter (optimized for maximum output power); right, 180\,µm beam (optimized for best beam quality).}
	\label{outputspectrum}
\end{figure}

The MIR output center wavelength can be controlled by simultaneous adjustments of (1) launch power to the highly-nonlinear fiber, (2) lateral position of the PPLN crystal and (3) pump-signal temporal overlap.
Figure \ref{outputspectrum}a shows output spectra for different center-wavelength settings together with achieved output power, and figure \ref{outputspectrum}b the obtained MIR beam profile under different focusing conditions. We choose the focusing conditions of signal- and pump- pulses for optimized MIR beam quality and low MIR divergence, resulting in output powers up to 300\,µW (compare Fig. \ref{outputspectrum}). The conversion efficiency peaks at 3.5\,µm center wavelength, since here the refractive indices of PPLN for signal and idler pulses are similar and temporal walk off is minimized.
The achieved power levels under this focusing conditions are sufficient to saturate the detectors of the Fourier transform spectrometer (FTS) used in our application for MIR absorption spectroscopy.
Tighter focusing leads to higher output powers at reduced beam quality, resulting in lower SNR in spectroscopy. For example, with focusing to approx. 60\,µm diameter it was possible to obtain output powers higher than 1\,mW at 4.6\,µm at a distorted beam profile (left profile in Fig. \ref{outputspectrum}b).

\section{DFG RIN measurements}
\begin{figure}[htbp]
	\centering
	\includegraphics[width=\linewidth]{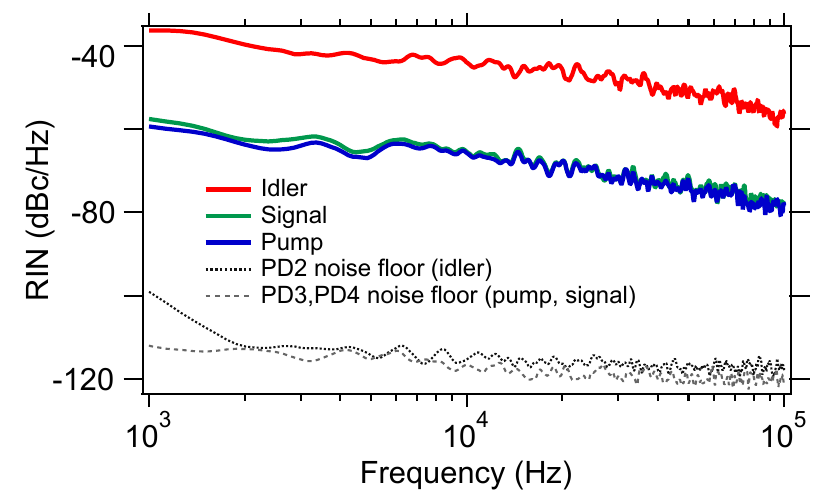}
	\caption{Relative intensity noise power spectral density of pump, signal and idler pulses.}
	\label{rinsignalidler}
\end{figure}

We observed that the relative intensity noise of our DFG source strongly depends on the temporal overlap of pump- and signal pulse in the nonlinear crystal. In a first experiment we characterized the source RIN power spectral density. The results are shown in Figure \ref{rinsignalidler}. For this measurement we detected pump- signal- and idler pulses after the PPLN crystal using photodetectors and a FFT analyzer \cite{scott2001high}. A germanium window and is a long pass filter (cut-on wavelength 1300\,nm) was used to separate idler,  pump- and signal pulses. The idler pulses were detected using a liquid nitrogen cooled InSb photodiode (size 0.5\,mm x 0.5\,mm, InfraRed Associates) connected to a transimpedance amplifier (bandwidth: 1.5\,Hz to 500\,kHz, InfraRed Associates).
Signal- and pump- pulses were characterized using InGaAs PIN photodiodes (diameter 1\,mm, Hamamatsu G12180-010A ) and transimpedance amplifiers, connected to the FFT analyzer.
For those measurements we manually adjusted pump-signal temporal overlap and acquired signal and idler RIN power spectral density simultaneously using two inputs of the FFT analyzer. We observed, that the pump and signal RIN power spectral densities are correlated and do not depend on pump-signal delay (compare Fig.3). We also observed a slight excess noise of about 1\,dB in the signal pulse RIN power spectral density, most likely caused by the supercontinuum generation process. As mentioned above, the RIN of the idler pulses was fluctuating strongly with changes on temporal overlap between signal- and pump- pulses. We were able to occasionally observe stable output parameters for a few seconds, allowing measurements of the idler RIN power spectral density. Fig. \ref{rinsignalidler} shows an example of such a short-term stable condition, most likely not under the optimum temporal overlap setting. One clearly observes excess noise of 23\,dB/Hz in average over the entire spectral range plotted in Fig. \ref{rinsignalidler}. For investigating the dependence of MIR RIN from temporal pump-signal pulse overlap, we modified our set-up using an integrating RIN detector: We electronically band-pass filtered the output of our photodetectors and connected the filter output to a demodulating logarithmic amplifier (Analog Devices AD8310) in a configuration for low-frequency signals (compare Ref. \cite{pilotte2005operation}).  Our set-up was configured such that the output of the integrating RIN detector was proportional to the logarithm of integral of the RIN power spectral density between 1\,kHz and 30\,kHz. With this detector integrated RIN fluctuations could be measured with a bandwidth of 1\,kHz and a dynamic range of more than 70\,dB.

\begin{figure}[htbp]
	\centering
	\includegraphics[width=\linewidth]{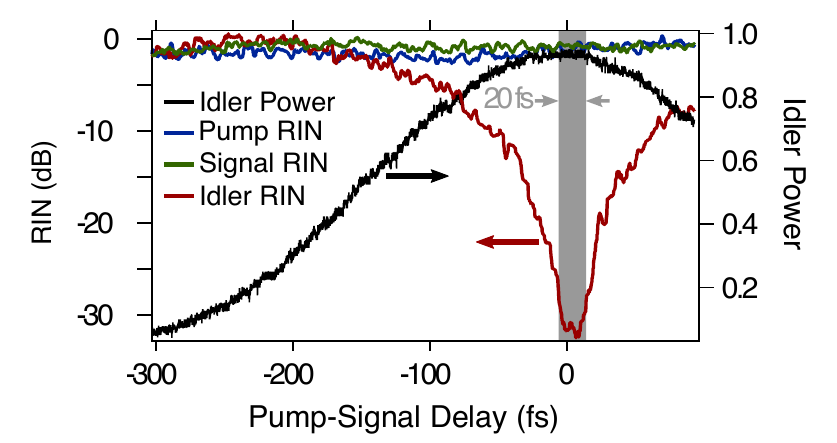}
	\caption{Integrated RIN power spectral density between 1 kHz and 30 kHz for pump, signal and idler pulses as a function of signal- pump temporal overlap (left axis) and idler power as a function of signal- pump temporal overlap (right axis). The shaded area represents the 20\,fs wide 3\,dB width of the idler RIN minimum. The idler power change in the same 20\,fs window is 1$\%$. At both edges of this region we observe an idler RIN slope of 20\,dB in 40\,fs pump-signal delay. All integrated RIN traces are offset such, that they overlap at -300\,fs. }
	\label{rinvsdelay}
\end{figure}

We now applied sinusoidal modulation (1\,Hz, 42\,µm amplitude) to a PZT transducer in the mechanical delay line which controls the temporal overlap between pump and signal pulses and recorded both the output of the integrating idler-RIN detector and the idler power simultaneously using an oscilloscope.
The result is shown in Fig. \ref{rinvsdelay} for signal- and idler- wavelengths of 1.36\,µm and of 4.6\,µm, respectively. Measurements at other center wavelengths setting resulted in very similar results: We observe a clear power maximum and RIN minimum for the idler output.

Remarkably, the 3 dB width of the RIN minimum corresponds to only 20\,fs pump-signal delay, while the idler-output power change for the same delay was about 1$\%$. We observed similar behavior at different focusing conditions, and after replacing the fan-out PPLN by a 1\,mm length PPLN with discrete poling periods, confirming that the observed behavior is not dependent on specifics of crystal or focusing conditions.

For our application of MIR frequency comb absorption spectroscopy both constant high signal-to-noise ratio of acquired FTS spectra and long-term stability of the DFG source are crucial. To achieve long-term stable, low RIN MIR output, we actively stabilized the signal- pump pulse temporal overlap such that the delay line was steered to the position for minimum idler RIN. We achieved this by implementing a dither-lock scheme by modulating the piezoelectric transducer in the delay line with 100\,nm amplitude (200\,nm of total path length, or approx. 0.6\,fs) at 200\,Hz and detecting the source RIN directly at the output of the InSb-detector of our FTS. To be able to detect simultaneously MIR power, MIR RIN and the FTS signal we electronically band-pass filtered the detector output and used the spectral ranges DC to 10\,Hz for MIR power measurement, 1\,kHz to 30\,kHz for RIN measurement and 100\,kHz to 500\,kHz for the FTS output signal. Note, the center frequency and bandwidth of the FTS output signal (interference fringes) is determined by the scan speed of the FTS spectrometer’s delay line and source bandwidth. We typically use  200\,kHz fringe center frequency. The error signal for feedback control was generated using a lock-in amplifier (SR530, Stanford Research Systems) and an analog PID controller (SIM960, Stanford Research Systems). Note that the 70\,dB dynamic range in the integrated RIN detection is orders of magnitude higher than the SNR of 300 on the spectrum, therefore no effect of modulation can be observed in the measured spectrum, even when averaging many spectra with a resulting SNR of 2000.

\section{Application to FTS spectroscopy}
\begin{figure}[htbp]
	\centering
	\includegraphics[width=\linewidth]{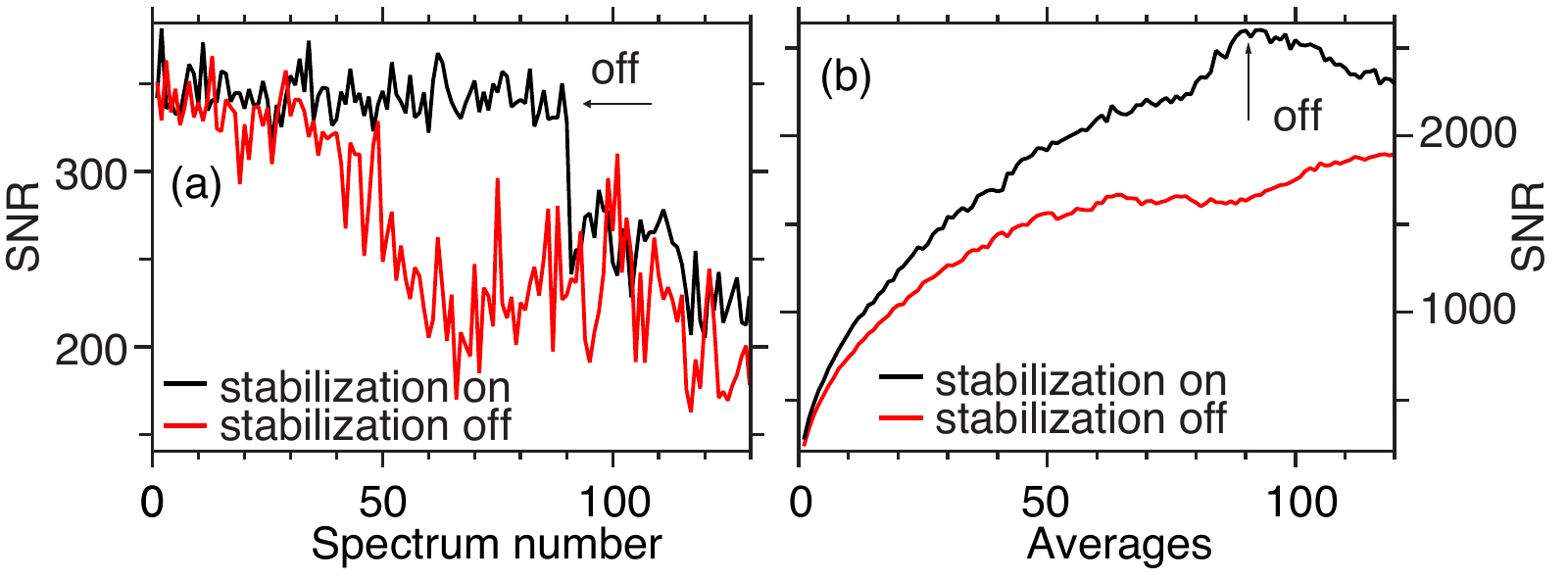}
	\caption{SNR of MIR spectra acquired by a home-built FTS spectrometer. (a) SNR (calculated via comparison of two subsequent spectra) as a function of acquisition number; (b) SNR calculated via comparison of the spectrum of the average of N interferograms to a single spectrum as a function of N.The arrows indicate the moment when the stabilization was turned off for the black trace.}
	\label{averaging}
\end{figure}

To test the performance of our active stabilization set-up we used the home-build FTS spectrometer for the MIR frequency comb spectroscopy application. We recorded a series of interferograms of our MIR source and calculated the corresponding spectra. We recorded the individual interferograms by scanning the FTS over 2\,m optical path delay in 9\,s intervals between scans. We calculated a signal to noise figure of merrit by dividing two consecutive spectra and calculating the ratio’s standard deviation in the FWHM bandwidth of our MIR source. This figure of merit corresponds to the signal-to-noise (SNR) ratio of absorption features with an absorption of 100$\%$ and will be referred to as SNR. Figure \ref{averaging} (left) shows the SNR for a sequence of measurements comparing consecutive spectra.
In a first test with stabilization off (red traces in Fig. \ref{averaging}) we observed in a well aligned delay setting a SNR of 300. After about 50 scans, the SNR decreased to 200 due to drifts in the system.
In a second test (black traces in Fig. \ref{averaging}) we show that the initial SNR of 300 can be kept until the stabilization was switched off after about 90 scans.  A constant SNR value over a large number of scans is extremely important when averaging is required. Figure \ref{averaging} (right) shows the results of a SNR calculation of the identical set of data, but now comparing the calculated spectrum from averaging N interferograms to a single spectrum as a function of N.  One can see in Fig \ref{rinvsdelay} right (black trace) that the averaged SNR improves from 300 (single spectrum) to 2600 after 80 scans corresponding to an improvement factor of 8.6, very close to the theoretically expected square-root of N value of 8.9. After switching off the stabilization, the SNR drops when more scans are added to the average. In a control experiment with stabilization off, only a SNR of 1900 could be reached, even when averaging over 120 scans. 

\section{Conclusion}
In conclusion we showed that the relative intensity noise power spectral density of a MIR DFG source can dramatically depend on the temporal overlap between pump- and signal- pulse. With active stabilization, we were able to operate over long times (up to several hours) our DFG MIR source at the lowest RIN. Without stabilization, the system drifted eventually to a state where up to 30\,dB higher RIN could be observed.

We would like to point out that the pump-signal path delay change corresponding to a transition from minimum to maximum integrated RIN corresponds to 51\,µm. This is considerably smaller than any other length and temporal scales involved in the experiment, namely  Rayleigh range (10\,mm), temporal walk-off length between pump and signal pulses (190\,µm) and PPLN crystal length (10\,mm or 1\,mm). 
The effect does not depend on crystal length or focusing conditions.  At this point we can only speculate about the explanation for this effect and think that accidental competing parametric processes could be one cause for excess noise.

The developed active stabilization method of our DFG MIR source was crucial for obtaining the required high stability and low noise level for molecular absorption spectroscopy in the MIR region. It allowed to maintain the MIR source stability over long times and to significantly  improve the SNR of measured spectra. It can improve also any other application based on light intensity measurement of DFG-based femtosecond sources. Our MIR source is now routinely used for absorption spectroscopy experiments.

\section*{Funding}
Vinicius Silva de Oliveira acknowledges the support of Conselho Nacional de Desenvolvimento Científico e Tecnológico (CNPq) - Brasil. 
Piotr Mas\l{}owski is supported by the Stephenson Distinguished Visitor Programme by DESY and the PIER Fellowship by DESY and the University of Hamburg, as well as by  National Science Center, Poland (project no 2016/23/B/ST2/00730).
\bibliography{biblioteca}

\bibliographystyle{abbrv}

\end{document}